\documentclass[
aps,
prl,
reprint,
superscriptaddress,
longbibliography,
amsmath,
amssymb,
floatfix.
xcolor=table]{revtex4-1}

\usepackage{graphicx}% Include figure files
\usepackage{dcolumn}% Align table columns on decimal point
\usepackage{bm}% bold math
\usepackage[utf8]{inputenc}
\usepackage[T1]{fontenc}
\usepackage{mathptmx}
\usepackage{subcaption}
\usepackage[mathlines]{lineno}
\usepackage{siunitx}
\usepackage{booktabs}
\usepackage[table,xcdraw]{xcolor}
\usepackage{multirow}
\usepackage{float}
\usepackage{hyperref}
\usepackage{color}
\usepackage{ulem}

\begin{document}
	\title[Tension-free Dirac strings and steered magnetic charges in 3D artificial spin ice]{Tension-free Dirac strings and steered magnetic charges in 3D artificial spin ice}% Force line breaks with \\
	
	\newcommand{\cn}{\textcolor{blue}}

	\newcommand{\univiefun}{Physics of Functional Materials, Faculty of Physics, University of Vienna, Waehringer Strasse 17, 1090 Vienna, Austria}
	\newcommand{\mmm}{University of Vienna Research Platform MMM Mathematics-Magnetism-Materials, University of Vienna, Austria}
	\newcommand{\la}{Theoretical Division, Los Alamos National Laboratory, Los Alamos, NM, 87545, USA}
	\newcommand{\nanomag}{Nanomagnetics and Magnonics, Faculty of Physics, University of Vienna, Boltzmanngasse 5, 1090 Vienna, Austria}

	\author{Sabri~Koraltan}
	\email{sabri.koraltan@univie.ac.at}

	\author{Florian~Slanovc}
	\author{Florian~Bruckner}
	\affiliation{\univiefun}
	
	\author{Cristiano~Nisoli}
	\affiliation{\la}
	
	\author{Andrii~V.~Chumak}	
	\author{Oleksandr~V.~Dobrovolskiy}
	\affiliation{\nanomag}
	
	\author{Claas~Abert}
    \author{Dieter~Suess}
	\affiliation{\univiefun}
	\affiliation{\mmm}

	\date{\today}
	\begin{abstract}
	
		3D nano-architectures present a new paradigm in modern condensed matter physics with numerous applications in photonics, biomedicine, and spintronics. They are promising for the realisation of 3D magnetic nano-networks for ultra-fast and low-energy data storage.  Frustration in these systems can lead to magnetic charges or magnetic monopoles, which can function as mobile, binary information carriers. However, Dirac strings in 2D artificial spin ices bind magnetic charges, while 3D dipolar counterparts require cryogenic temperatures for their stability. Here, we  present a micromagnetic study of a highly-frustrated 3D artificial spin ice harboring tension-free Dirac strings with unbound magnetic charges at room temperature. We use micromagnetic simulations to demonstrate that the mobility threshold for magnetic charges is by $\SI{2}{eV}$ lower than their unbinding energy. By applying global magnetic fields, we steer magnetic charges in a given direction omitting unintended switchings. The introduced system paves a way towards 3D magnetic networks for data transport and storage.
		\end{abstract}
	\maketitle
	
	\section{Introduction}
	Data storage and transport devices ranging from hard disk drives to flash memories, from CMOS to spintronic technologies are of crucial importance in today's technological world. 
	Usually, these devices are based on 2D structures approaching their limitations every day. 3D structures started to emerge over the past years, leading to significant improvements in both reducing the dimensions and increasing their efficiency, e.g. flash memories~\cite{zhu_ren_song_2019, dieny_opportunities_2020}. In these data storage devices, the third dimension is used by the simple stacking of the same 2D structures. Thus, the whole power of the additional third dimension is not utilised.
	
	Over the past years, spin ices, a class of 3D materials, have been investigated in detail. Spin ices are frustrated systems where the magnetic moments are residing on the sites of a pyrochlore lattice, a lattice with corner sharing tetrahedra, and commonly referred to as dipolar spin ices~(DSI) \cite{harris_geometrical_1997, bramwell_frustration_1998, ryzhkin_magnetic_2005, castelnovo_magnetic_2008, morris_dirac_2009, fennell_magnetic_2009, bramwell_history_2020}.
    Its degenerate ground state obeys the \textit{ice-rule}, where two magnetic moments point to the center of the tetrahedra and two away from it.
    Switching a magnetic moment breaks the ice rule and  creates a pair of magnetic charge in the centers of the vertices \cite{ryzhkin_magnetic_2005, castelnovo_magnetic_2008}. These magnetic charges, commonly referred to as \textit{emergent magnetic monopoles}, can be separated with a finite energy cost, and propagated through the lattice. In a classical analogue to Dirac's  theory of magnetic monopoles \cite{dirac}, monopole motion in spin ice leaves a trace called a Dirac String~(DS), which is simply the chain of flipped magnetic moments connecting the two separated positive and negative magnetic poles. 
	Theoretical and experimental studies have shown that the energetic ground state is degenerate, constrained by the ice rule and that the magnetic monopoles are connected via Dirac Strings at low temperatures \cite{ryzhkin_magnetic_2005, castelnovo_magnetic_2008, harris_geometrical_1997, bramwell_frustration_1998, morris_dirac_2009, fennell_magnetic_2009, harris_geometrical_1997, matsuhira_new_2002, ryzhkin_dynamic_2013}.
	
	In order to study the geometrical frustrations and the magnetic charges on more controllable platform, 2D artificial spin ices~(2DASIs) have been designed and investigated in detail~\cite{wang_artificial_2006, moller_artificial_2006, mol_magnetic_2009, mol_conditions_2010, nisoli_colloquium_2013, skjaervo_advances_2020}. There, lithographically patterned nanomagnets are arranged on different lattices. The most common ASI lattices are the square~\cite{nisoli_ground_2007, kapaklis_melting_2012, kapaklis_thermal_2014, farhan_direct_2013} and Kagome ices~\cite{farhan_magnetic_2017, qi_direct_2008, arnalds_thermalized_2012, hofhuis_thermally_2020}.
	
	In contrast to DSI,  the reduced dimensionality and geometric frustration in square ice  lifts the degeneracy of the ice rule~\cite{nisoli_colloquium_2013, skjaervo_advances_2020, chern_degeneracy_2013, moller_artificial_2006, mol_magnetic_2009, mol_conditions_2010,nisoli_equilibrium_2020}, thus limiting monopole mobility. Various methods have been explored to regain spin ice degeneracy, including quasi three-dimensional lattices \cite{mol_conditions_2010,moller_artificial_2006, farhan_emergent_2019, chern_realizing_2014, thonig_thermal_2014, perrin_extensive_2016, farhan_geometrical_2020, mistonov_magnetic_2019} and interaction modifiers~\cite{ostman2018interaction}. After early designs~\cite{chern_realizing_2014} and  realizations~\cite{mistonov2013three} had envisioned a 3D artificial spin ice, the first three dimensional frustrated nanowire-lattice~\cite{may_realisation_2019} was manufactured by two-photon lithography~\cite{williams_two-photon_2018}, in which charge propagation was later demonstrated~\cite{may_magnetic_2020}. However, in this lattice, the degeneracy of the ground state is still lifted, as the 3D structure consists of an interconnected nanowire-lattice. Additionally, the magnetic charges are connected via DS's, which store energy due to the presence of domain-walls at the vertex centers.
	
	In this work, we combine the advantages of both DSI and ASI and present a 3D artificial spin ice~(3DASI lattice), where numerical investigations are performed. Our lattice is a 3D nano-magnetic network of vertices consisting of four disconnected 3D magnetic ellipsoids with perfect Ising behavior enabling tension-free DS's and unbound magnetic charges by recovering the lost degeneracy of the ice rule obeying states.   
	In this new lattice, we conserve the full accessibility of 2DASI and investigate numerically the propagation of emergent magnetic monopoles. We demonstrate that the difference in energy, required to create magnetic charges and their transport, is around $\SI{2}{eV}$, which enables the controlled propagation of unbound charges. 
	Considering the emergent magnetic monopoles as binary, mobile information carriers, the presented lattice demonstrates the steered motion of charge carriers in a 3D magnetic nano-network at room temperatures. The controllability of these charges paves a way towards a 3D magnetic network for data transport and storage.
	
	\section{Results}
	\textbf{Modeling.} In recent years, the direct-write technique of focused electron beam induced deposition~(FEBID) has reached a high level of maturity for the 3D nanofabrication. Many  complex-shaped nano-architectures have become available, providing access to experimental investigations of curvature-, geometry- and topology-induced effects in various disciplines, including magnetism, superconductivity, photonics and plasmonics \cite{fernandez-pacheco_writing_2020, keller_direct-write_2018, dobrovolskiy_spin-wave_2020, huth_focused_2018, fernandez-pacheco_three-dimensional_2017}.
	
\begin{figure}[t]
	    \includegraphics[width=0.95\columnwidth]{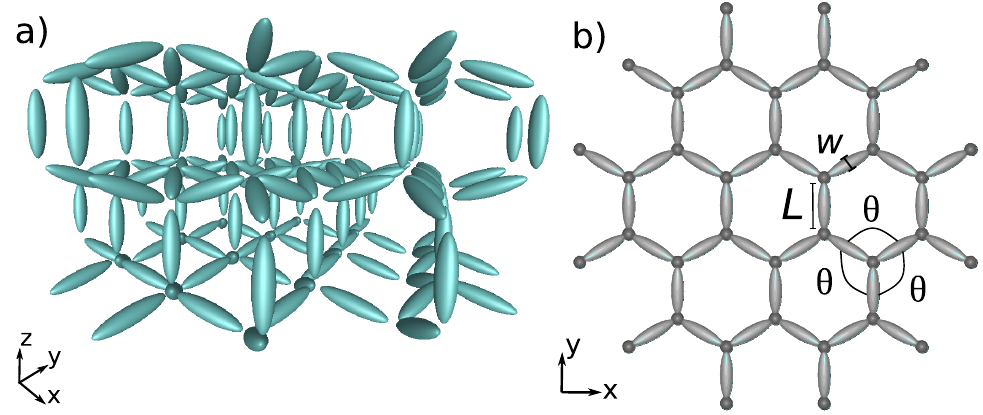}
	    \caption{\textbf{3DASI Lattice.}(a) 3D Illustration of the 3DASI lattice, and (b) the top view. The length of the ellipsoids $L$ and maximal width $W$ as well as the angle between the ellipsoids $\theta$ are depicted in (b).}
	    \label{fig:lattice}
	\end{figure}

	Inspired by the recent developments in 3D optical lithography and focused particle 3D nano-printing by FEBID, we present a three dimensional ASI~(3DASI) lattice, where magnetic rotational ellipsoids are arranged along the main axis of a tetrahedron, resulting in an angle $\theta = \mathrm{\arccos(-1/3)} \approx \SI{109.5}{^\circ}$ between the elements, reproducing the Ice $\text{I}_h$ crystal of the water ice \cite{harris_geometrical_1997, ryzhkin_magnetic_2005, bramwell_history_2020}. Figure~\ref{fig:lattice} illustrates the designed 3DASI lattice. 
	
	Note that in Fig.~\ref{fig:lattice} we illustrate only the magnetic ellipsoids forming the lattice. In reality, the ellipsoids can be fabricated by direct-writing, i.e. FEBID, and interconnected with a magnetic insulator, e.g. platinum-~\cite{keller_direct-write_2018} or niobium-based~\cite{porrati_crystalline_2019} compounds. We provide the illustration of a rather realistic fabrication model for one single vertex in the supplemental materials.
	With FEBID as a suitable nanofabrication technique of the 3DASI lattice, maximal width of the ellipsoids $w$ can be reduced down to few tens of nanometers, while the length $L$ can be chosen up to a few micrometers \cite{huth_focused_2018, fernandez-pacheco_writing_2020, dobrovolskiy_spin-wave_2020, keller_direct-write_2018}.

	We choose rotational magnetic ellipsoids as ASI elements, because the self-demagnetizing field of the ellipsoids is homogeneous thus keeping the magnetic elements uniformly magnetized in the direction of the longer axis with no edge inhomogeneities~\cite{osborn_demagnetizing_1945, rougemaille_chiral_2013, koraltan_dependence_2020, leo_chiral_2020}. Hence, our model shows nearly perfect Ising behavior and is suitable to separate and host magnetic charges. The maximal average deviation from a perfect uniformly magnetization for one ellipsoid in our model is found to be $\Delta\phi \approx \SI{4e-4}{deg}$, where the interactions with nearest and next-nearest neighbors were encountered. 
	
	In this work, we restrict our model to three layers to conduct numerical experiments in feasible times. However, we use the top and bottom layer rather as boundary conditions, while the region of interest is only the middle layer. In other words, we analyze the 3DASI lattice as a bulk system.

	\textbf{Dirac Strings and magnetic monopoles.} Because of the symmetry of the tetrahedra, the degeneracy of the ice rule is not lifted in 3DASI. Consequently, the tension of the DS should vanish. To obtain an energy scale, we approximate the ellipsoids by magnetic dipoles with the dipole moment $\mu = M_s V$, where $M_s$ is the saturation magnetization and $V$ the volume of the nanostructure. In this formulation, the dipolar interaction energy for each pair of magnetic dipole moments $\mathbf{m}_{i},\mathbf{m}_{j}$ at positions $\mathbf{r}_{i}, \mathbf{r}_{j}$ in the lattice is given by the dipolar interaction energy
    \begin{equation}
        E_{\mathrm{dip}}=\frac{\mu_{0}}{4 \pi\left|\mathbf{r}_{ij}\right|^{3}}\left[\mathbf{m}_{i} \cdot \mathbf{m}_{j}-3 \frac{\left(\mathbf{m}_{i} \cdot \mathbf{r}_{ij}\right)\left(\mathbf{m}_{j} \cdot \mathbf{r}_{ij}\right)}{\left|\mathbf{r}_{i j}\right|^{2}}\right]
        \label{eq:e_dip}
    \end{equation}
    with $\mathbf{r}_{ij} := \mathbf{r}_{j} - \mathbf{r}_{i}$ and vacuum permeability $\mu_0$.
    Then from Eq.~\eqref{eq:e_dip} we obtain the energy scale
    \begin{equation}
        J_{NN} = \frac{3}{2\sqrt{2}}\frac{\mu_0}{\pi} \frac{\mu^2}{a^3}
    \end{equation}
    as introduced in \cite{leo_chiral_2020} , where $a$ is the lattice constant.
	
    \begin{table}[]
        \centering
        \begin{tabular}{m{0.23\columnwidth}m{0.35\columnwidth}m{0.35\columnwidth}}
        \toprule
        \toprule
        Vertex type & $E^{2D}_{dip} \mathrm{(J_{NN})}$ & $E^{3D}_{dip} \mathrm{(J_{NN})}$  \\
        \midrule
        Type I  & $2(\sqrt{2}/3 -2)$ & \multirow{ 2}{*}{$-5/(2\sqrt{3})$}  \\
        \cmidrule{1-2}
        Type II  & $-2\sqrt{2}/3$ &   \\
        \midrule
        Type III  & $0$ & $0$ \\
        \midrule
        Type IV  & $2(\sqrt{2}/3+2)$ & $15/(2\sqrt{3})$\\
        \bottomrule
        \bottomrule
        \end{tabular}
        \caption{Dipolar interaction energies in each vertex type in 2D and 3D ASI in units of $J_{NN}$, which is $J_{NN} \approx  \SI{1.72}{eV}$ for the chosen material and geometry parameters.}
        \label{tab:vertex}
    \end{table}
    
    From it we obtain the energy levels for the different vertex types, shown in Table~\ref{tab:vertex}. Note the degeneracy in the ice rule obeying vertices.  
	\begin{figure*}[t]
		\includegraphics[width=0.95\linewidth]{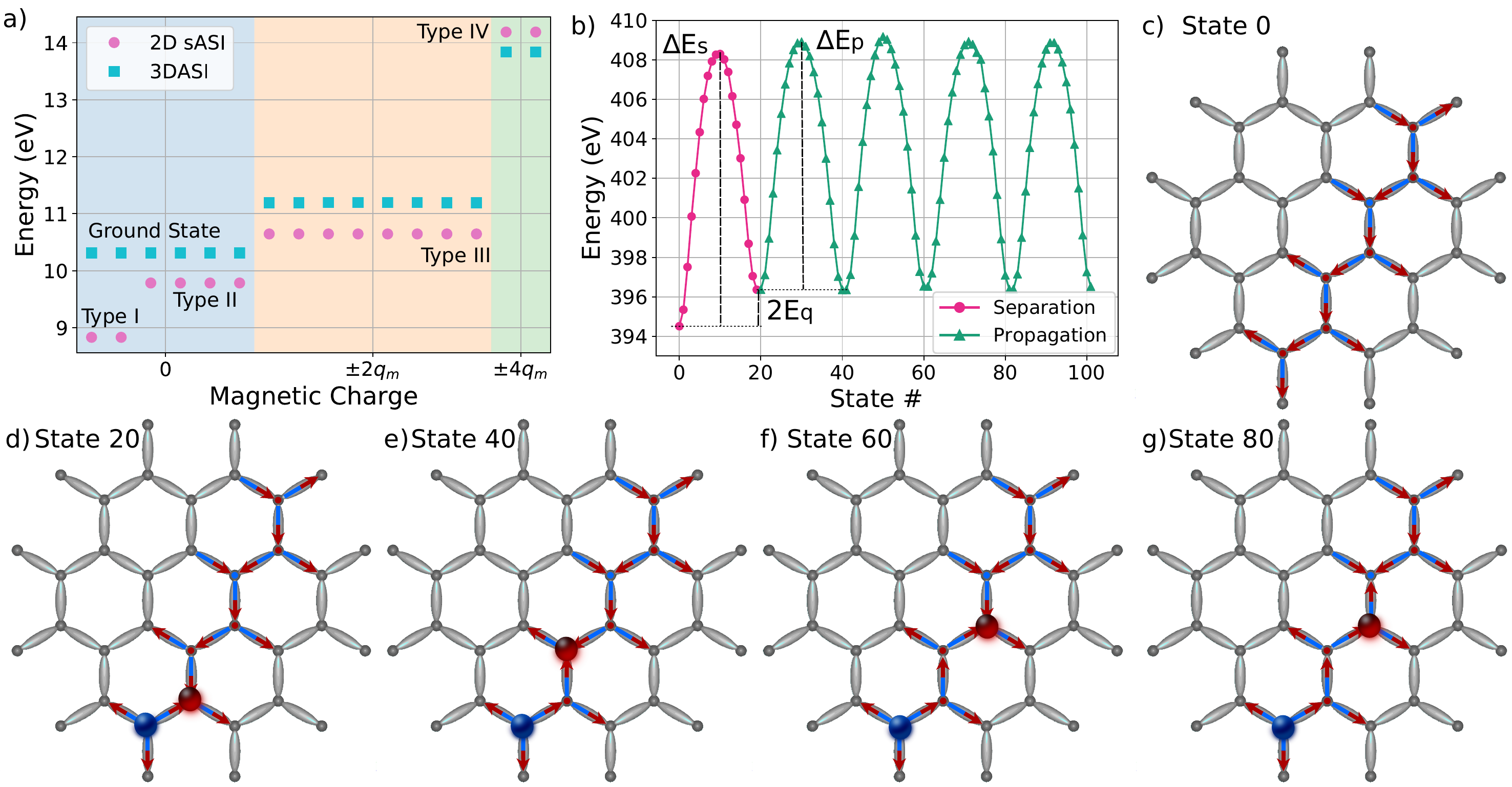}
		\caption{\textbf{Magnetostatic energies and minimum energy paths for propagation and separation of charges.}(a) Total energies per vertex calculated with a micromagnetic treatment by taking into account energetic contributions of the exchange and demagnetization fields, where every possible magnetic configurations in a 2DsASI (circles) and in our 3DASI~(squares) was considered.  (b) Minimum energy paths to separate (magenta) the two magnetic charges by switching  one ellipsoids magnetization in the ground state, and to propagate it further through the lattice in a constant direction (green). The difference of saddle points and the corresponding minima yields the separation barrier $\Delta E_s$ and propagation barrier $\Delta E_p$, while energy of the lattice increases by $2E_q \approx \SI{2}{eV}$ due the two additional charge defects. (c)~-~(g) Schematic illustrations of the magnetization configurations for the ellipsoids in the Ds with the positive (dark red) and negative (dark blue) charged magnetic monopoles. Snapshots of magnetizations and animations available in the supplemental materials.}
		\label{fig:v_en}
	\end{figure*}
	
    Additionally, we perform calculations via micromagnetic simulations using \texttt{magnum.fe}\cite{abert_magnumfe_2013} using material properties similar to the Cobalt-Iron($\mathrm{CoFe}$) alloys used in FEBID. Note that with this technique, the material parameters are already tunable~\cite{dobrovolskiy_spin-wave_2020, bunyaev_engineered_2020} for 2D structures, and advances are expected in the near future. We choose a saturation magnetization $M_s = \SI{800}{kA/m}$ and exchange stiffness constant $A_{\mathrm{ex}}=\SI{15.5}{pJ/m}$. The ellipsoids have the dimensions $L=\SI{100}{nm}$, $w=\SI{20}{nm}$ and a vertex-to-vertex distance of $a=\SI{120}{nm}$. It is noteworthy that in our studies we especially focused on the 3D structures of sizes that have been demonstrated for platinum based nanofabrications. Magnetic CoFe/CoFeB structures are larger, but the down-scaling is expected within the next years. 
    
    Figure~\ref{fig:v_en}a) shows the calculated energies of all the possible vertices in 2DsASI and 3DASI. The latter show clearly the 6-fold degeneracy of the ground state Type I/II. This degeneracy allows for magnetic charges to be separated, because a tension-free DS is created, making them \textit{truly} unbound. 
    In Fig.~\ref{fig:v_en}a), we chose to plot all $2^4=16$ possible configurations within a vertex. In reality, for vertices in the 3DASI lattice there exists only one unique ground state configuration, the ice-rule configuration, up to symmetry operations, i.e. rotations and mirroring, always obeying the \textit{2in-2out} ice rule.
    
    A more detailed description of the simulations, geometries and additional snapshots of magnetizations are given in the supplemental material. 

	According to the original definition of emergent magnetic monopoles, the energy required to create a monopole-antimonopole pair~(charge separation), should be larger than the energy required to propagate it~\cite{dirac, vedmedenko_dynamics_2016, castelnovo_magnetic_2008, mol_conditions_2010, jaubert_signature_2009, jaubert_magnetic_2011}.
    We verify these properties by applying a full-micromagnetic model to calculate the energy barriers to separate and propagate the magnetic charges\cite{e_simplified_2007, abert_micromagnetics_2019, koraltan_dependence_2020}. The detailed description of the full micromagnetic model and its application on 2DASIs is given in Ref.~\cite{koraltan_dependence_2020}.
    
    Figure~\ref{fig:v_en}b) illustrates the minimum energy paths~(MEPs) for the switching processes within a DS. Starting from a lattice in the ground state, as illustrated in Fig.~\ref{fig:v_en}c), we separate two magnetic charges by switching the magnetization of one ellipsoid, and thus creating two Type~III defects, which are depicted in Fig.~\ref{fig:v_en}d). The positive charge is then propagated to increase the length of the DS, shown in Figs.~\ref{fig:v_en}e)-g). It can be seen, that the separation MEP (magenta) yields an separation barrier $\Delta E_s$ approximately $\SI{2}{eV}$ higher than the propagation barriers $\Delta E_p$ (green MEPs), which are nearly constant. The reversal mechanisms are highly non-coherent and complex, being dictated by domain wall movements. 
    
    By separating the emergent monopoles, the energy of the system increases about $2E_q = \SI{2}{eV}$, which represents the energy increase due to Type~III defects. Since we propagate only one charge, while the environment does not change, the energy of the system should stay constant during the propagation. However, we need to consider the possible Coulomb interactions between magnetic charges~\cite{ryzhkin_magnetic_2005, castelnovo_magnetic_2008}. In contrast to the classical DSI, this interaction can be neglected in comparison to the energy barriers in our 3DASI lattice~\cite{castelnovo_magnetic_2008}.

    \begin{figure*}[]
		\includegraphics[width=\linewidth]{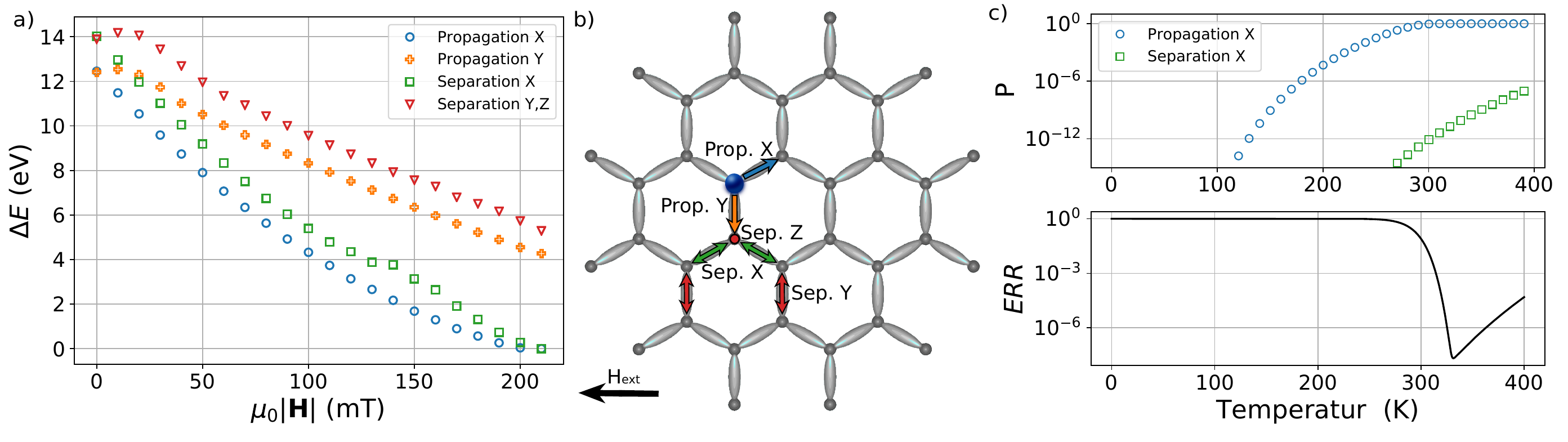}
		\caption{\textbf{Arbitrary field finite temperature micromagnetic analysis.} a) Dependence of the energy barrier on the applied external fields at $T=\SI{0}{K}$ to switch the magnetizations of different ellipsoids as depicted in (b). c) Upper panel: the switching probability P at different temperatures for the relevant transitions from (b) calculated with FTM-analysis, Eq.~\eqref{eq:prob}, where a Zeeman field with $\mu_0|\mathbf{H}| =\SI{180}{mT}$, attempt frequency $f_0 = \SI{50}{GHz}$ and pulse time $t_{\mathrm{final}} = \SI{0.15}{ns}$ were chosen. Lower panel: mean error rate of successful propagation correlated with the unintended separation, as given by Eq.~\eqref{eq:err}}
		\label{fig:ftm}
	\end{figure*}

    Note that these results are obtained at $T=\SI{0}{K}$. In order to analyze the dependence of the stability and the propagation properties of a single monopole on both externally applied fields, and temperature, we use the arbitrary field finite temperature micromagnetic analysis~(FTM)~\cite{dieter_calculation_2011}.

    \textbf{Arbitrary field finite temperature micromagnetic analysis.} Starting from an initial magnetization state with one magnetic monopole of charge $Q=\SI{-2}{q_m}$, we calculate the energy barriers to switch the magnetization of different 3DASI elements under a  uniform, external magnetic field with $\mu_0\mathbf{H}=-\mu_0|\mathbf{H}|\mathbf{e_x}$.
    Here, $\mu_0|\mathbf{H}|$ denotes the strength of the applied field, and $\mathbf{e_x}$ is the unit vector along $x$. $\mu_0|\mathbf{H}|$ is varied between $0-\SI{200}{mT}$.
    
\begin{figure*}[t!]
		\includegraphics[width=\linewidth]{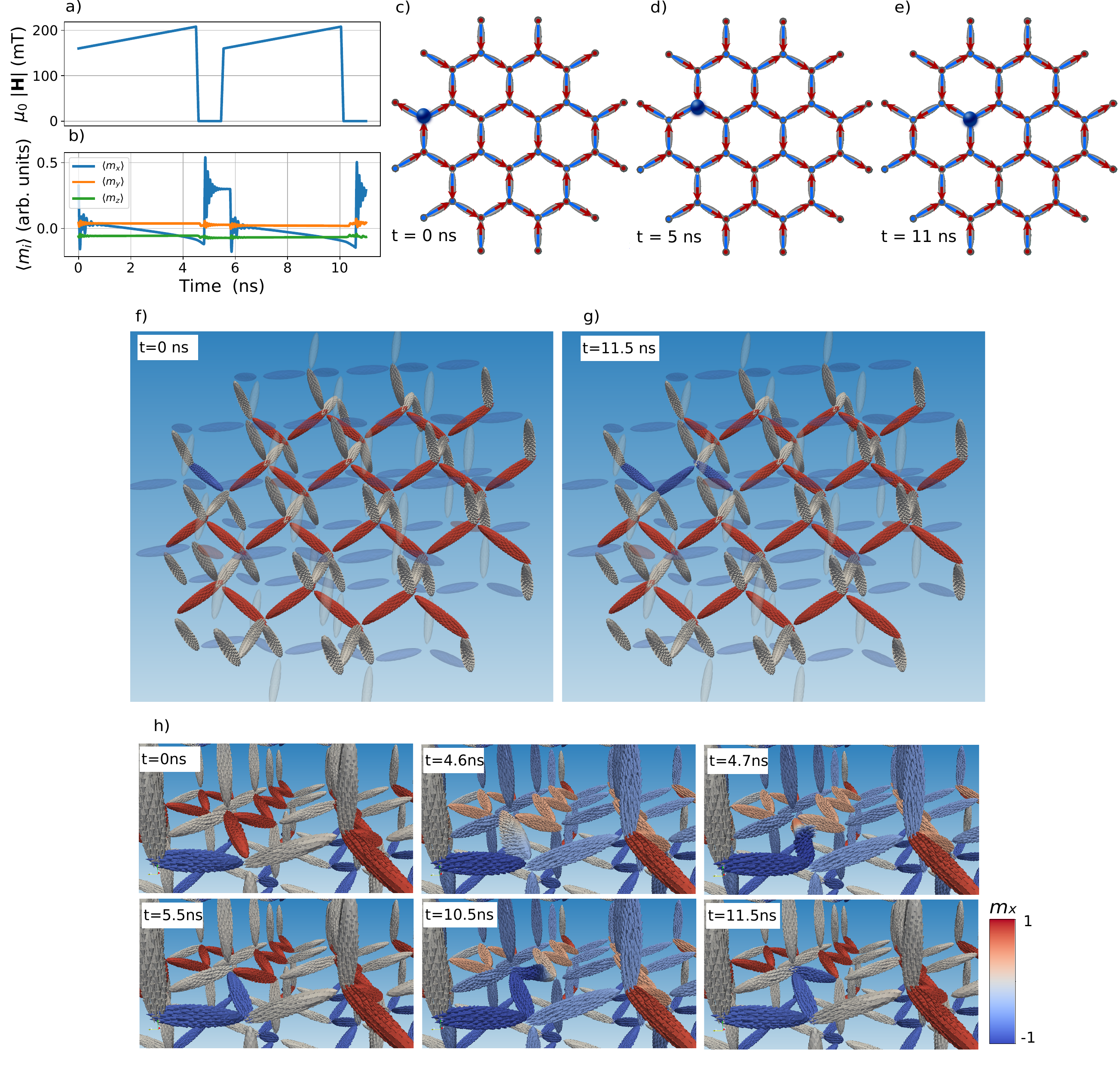}
		\caption{\textbf{Field-driven monopole propagation.} Temporal evolution of the magnitude of the applied field (a) and of the averaged magnetization components (b) within the bulk layer during the x-propagation. The field is applied along $-x$ in order to maximize the Zeeman energy acting on the ellipsoid whose magnetization should be switched. The field is turned off after $t=\SI{4.8}{ns}$ in order to transfer the magnetic charge to the next vertex controlling the propagation. (c)-(e) Schematic illustration of the monopole~(dark blue) propagation. (f)-(h) Snapshots of the magnetization states from the output of the conducted micromagnetic simulations, where f) illustrates the initial and g) the final magnetic configurations. The magnetization evolution during the simulations is depicted at different times (white boxes) in h). Color in f)-h) shows the $x$-component of the magnetization, as given by the colorbar.}
		\label{fig:m_vs_t}
	\end{figure*}

    The initial magnetization state is the same as in Fig.~\ref{fig:m_vs_t}e). We consider different transitions, where the magnetic charge is propagated along $x$ (blue arrow) and $y$ (orange arrow), or additional charges are separated in a neighbored vertex, where the separation in three different directions is considered, $x$ (green arrows), $y$ and $z$ (red), as illustrated in Fig.~\ref{fig:ftm}b). By doing so, we cover all possible transitions in our lattice, where the existing charges with $Q=\SI{-2}{q_m}$ are either propagated or new monopole-antimonopole pairs are separated at different locations.
    
    Figure~\ref{fig:ftm}a) shows the calculated energy barriers as a function of the applied field strength for the transitions depicted in Fig.~\ref{fig:ftm}b). Our results indicate that the propagation barrier along $X$ (blue, circles) is always lower than any other barrier. At vanishing fields, the barriers for propagation $X$ and $Y$ are equal, as demonstrated above. With the increasing field and approaching the coercive field, all barriers are lowering, ultimately, vanishing once the critical field is reached.   
    
    In the FTM analysis, described by Suess and co-workers in Ref.~\cite{dieter_calculation_2011}, a field-driven transition occurs with the switching probability
    \begin{equation}
        P=1-\mathrm{exp}\left( -f_0t_\mathrm{final}e^{\left(\dfrac{-\Delta E(H)}{k_BT}\right)} \right),
        \label{eq:prob}
    \end{equation}
    where $f_0$ denotes the attempt frequency, $\Delta E(H)$ the energy barrier at a given field magnitude, $k_B$ the Boltzmann constant, $t_\mathrm{final}$ the time of applied field pulse and $T$ is the temperature. 
    
    To analyze the stability of the propagation at room temperature, we consider $T=\SI{300}{K}$ and calculate the associated probabilities where we choose $\mu_0|\mathbf{H}| = \SI{180}{mT}$, $f_0 = \SI{50}{GHz}$ and $t_\mathrm{final}=\SI{0.15}{s}$. $f_0$ is chosen according to values considered in ASI literature~\cite{farhan_exploring_2013}, and $t_\mathrm{final}$ based on experimental expectations. Note that these two values appear only as a common factor in Eq.~\eqref{eq:prob} and therefore other values for $f_0$ can always be compensated by an appropriate choice of $t_\mathrm{final}$. The field strength $\mu_0|\mathbf{H}| = \SI{180}{mT}$ is chosen such that the switching barrier of the propagation is below 300$k_BT$. 
    
    The upper panel of Fig.~\ref{fig:ftm}c) illustrates the relevant probabilities for the switching possibilities depicted in Fig.~\ref{fig:ftm}b). Our results indicate that only the element of interest will change its magnetization, as its switching probability increases with the temperature, where $P_\mathrm{prop}(\SI{300}{K}) > 0.9$. All the other probabilities are nearly zero at temperatures around $T=\SI{300}{K}$. We chose to depict only the switching probabilities for the propagation and separation along $x$, as the other probabilities are lower than $\num{1e-15}$.
    
    However, we still need to define the mean switching error rate 
    \begin{equation}
        \mathrm{ERR}(T) = 1-P_\mathrm{prop}\left( 1 - P_\mathrm{sep}(T)\right)^N,
        \label{eq:err}
    \end{equation}
    which describes the probability for an unsuccessful propagation, i.e. falsely switch of one element leading to a separation (or propagation failure due to insufficient thermal energy) in a lattice with $N$ elements, where $P_\mathrm{prob}$ describes the propagation probabilities and $P_\mathrm{sep}$ the separation probabilities.
    
    In the lower panel of Fig.~\ref{fig:ftm}~c) we see that the error rate significantly decreases after $T=\SI{290}{K}$, as the $X$ propagation probability increases to nearly 1. For $T>\SI{330}{K}$ the propagation is guaranteed, however the separation probability increases leading to an increment in the mean error rate. 
    This indicates that the monopoles can be propagated in a controlled manner, in the desired direction, and that they can be accessed at temperatures around room temperature, improving the scalability and thermal stability of pyrochlore DSI.
    
    \textbf{Field-driven controlled monopole propagation.} Nevertheless, the controlled propagation can be achieved at any temperature, including $T=\SI{0}{K}$, if one adjusts the strength, direction and duration of the globally applied magnetic fields. As illustrated in Fig.~\ref{fig:ftm}-a), the energy barrier vanishes for large enough fields.
    To investigate this in a rather direct manner, we conduct numerical experiments, where we continue with our micromagnetic treatment and investigate the propagation of monopoles by applying pulses of external magnetic fields at $T=\SI{0}{K}$. 
    
    We start from an initial system where the middle layer is in the ground state.
    Reversing the magnetization in one boundary ellipsoid hosts exact one magnetic charge with $Q = -2q_m$, see Figs.~\ref{fig:m_vs_t}c) and f). We continue by applying a magnetic field along $-x$ direction, as in the FTM analysis above. The field profile can be seen in Fig.~\ref{fig:m_vs_t}a), where the magnitude of the applied field increases linearly over $\SI{4.8}{ns}$ from $160-\SI{208}{mT}$, and is then turned off abruptly after the propagation is successful. Note that we are confining the 3DASI lattice to avoid finite size effects, and analyze only the middle \textit{bulk} layer.

    Figure~\ref{fig:m_vs_t}b) shows the temporal evolution of the averaged magnetization components within the middle layer. 
    Our results indicate that the magnetic charge is propagated to the next vertex via domain wall movements as predicted by our string method calculations and FTM analysis. Magnetization snapshots at chosen times from Fig.~\ref{fig:m_vs_t}h) shows that until the critical field is reached, small deviations from the ising state are observed for the ellipsoids in the ground state, explaining the slow decrease in the average $x-$component of the magnetization from Fig.~\ref{fig:m_vs_t}b). Once the applied field is large enough, the propagation takes place keeping the other vertices in the ground state, avoiding additional charge separations. By turning off the external field, we relax all the ellipsoids back to their ising state and control the propagation by demand. One can see in Fig.~\ref{fig:m_vs_t}b) that the average magnetization drops rather abruptly when the magnetic charge is propagated one vertex further with each step, demonstrating the free and controlled propagation of an unbound emergent charge in the presented lattice.
    In the supplemental materials we also demonstrate that the magnetic charge can be further propagated to the next $x$-row by applying the same protocol with a magnetic field along $y$. 

    \section{Discussion}
    We investigated micromagnetically a new 3DASI lattice, which combines the advantages of classical 2DASI and pyrochlore DSI lattices, recovering the lost frustration and degeneracy of the ground state of the 2DsASI by enabling tension-free Dirac Strings and thermally stable unbound magnetic monopoles. Due to large distances compared to the pyrochlore DSI, the Coulomb interactions between the charges are negligible. String method simulations and finite temperature micromagnetic analysis show that the lattice allows to create and propagate magnetic charges, which can exist freely or in pairs being connected via tension-free Dirac strings at arbitrary temperatures. We showed by numerical experiments, that a single unbound charge can be propagated through the lattice, without creating additional magnetic charges.  
    
    Low-energy and ultra-fast switching in the 3DASI lattice, as well as the stability of emergent magnetic monopoles at room temperature and above, paves a way towards functional 3D magnetic nano-networks for data transport and storage. The controlled propagation of the magnetic charges is of high interest regarding the idea of \textit{magnetricity}~\cite{vedmedenko_dynamics_2016, arava}, which might lead to further new devices\cite{skjaervo_advances_2020}.
    
    Even though we focused only on the center layer and therefore regarded our model as a bulk lattice, the limitations of the $z$ layers, might show additional and equally interesting phenomena which will be investigated in detail in a further work. By \textit{shaving off} at a given $z$ dimension, the lattice would have magnetic vertices, containing only three ellipsoids, which would \textit{always} host a magnetic charge. The top and bottom layers could act as charge plasma layers, which might \textit{decharge} in the middle layer, creating a magnetic capacitor.

    \section*{Methods}
    \textbf{String Method.} For each switching process, we start from an initial state and a final state interconnected via 19-20 magnetization states, corresponding to a coherent rotation of the magnetization of the element of interest. In a second step, the total energy of each image is driven a small step towards its energetic minimum. The energy contributions are the demagnetization energy, the exchange energy and Zeeman energy. The third step consists in cubic interpolation of the new path, such that the magnetization states are equidistant according to an appropriate energy norm~\cite{e_simplified_2007, abert_magnumfe_2013, koraltan_dependence_2020}. The last two steps are repeated until convergence is reached.\\
    \textbf{Micromagnetic Simulations.} By using the finite and boundary element method based simulation code \texttt{magnum.fe}~\cite{abert_magnumfe_2013}, we solve the Landau-Lifshitz-Gilbert (LLG) equation~\cite{abert_micromagnetics_2019}:
    \begin{equation}
        \dfrac{\partial\mathbf{m}}{\partial t}=-\dfrac{\gamma}{1+\alpha^2}\mathbf{m}\times\mathbf{H^\mathrm{eff}}-\dfrac{\alpha\gamma}{1+\alpha^2}\mathbf{m}\times \left(\mathbf{m}\times\mathbf{H^\mathrm{eff}}\right),
        \label{eq:LLG}
    \end{equation}
    where $\alpha$ is the Gilbert damping constant, $\gamma$ the reduced gyromagnetic ratio, $\boldsymbol{m}$ the magnetization unit vector, and $\mathbf{H^{\mathrm{eff}}}$ is the effective field term, which includes the considered energy contributions from demagnetizing, exchange, external and anisotropy fields~(magnetic confinement). Vertex state energies are computed by calculating the demagnetizing and exchange energies from the relaxed structures~(high damping $\alpha=1$).
    For the field-driven dynamic simulations, we also include global external fields, and solve the LLG for a given time. Additionally, the magnetization components are averaged within the middle~(bulk) layer for each time step, as depicted in Fig.~\ref{fig:m_vs_t}b).
    \section*{acknowledgements}
	We would like to thank Kevin Hofhuis and Johann Fischbacher for the fruitful discussions. The computational results presented have been achieved, in part, using the Vienna Scientific Cluster (VSC).
	S.K., C.A. A.V.C. and D.S. gratefully acknowledge the Austrian Science Fund (FWF) for support through Grant No. I 4917 (MagFunc).
	O.V.D. acknowledges the Austrian Science Fund (FWF) for support through Grant No. I 4889 (CurviMag).
	
	\section*{Author contributions}
    S.K. and F.S. conceived the concept.
    C.A. and F.B. wrote and improved the micromagnetic codes.
    S.K performed the micromagnetic simulations.
    S.K., F.S., F.B., C.N., A.V.C., O.V.D., C.A and D.S. further improved and developed the concept and discussed the results.
    All authors contributed to the manuscript.

\section*{Competing interests}
The authors declare no competing interests.

\section*{Additional information}
\textbf{Supplementary information} is available for this paper at .

\textbf{Reprints and permission information} is available at .

Correspondence and requests for materials should be addressed to S.K.

\bibliography{newbib.bib}
\end{document}